Transport properties of the new Fe-based superconductor $K_xFe_2Se_2$ ($T_c$ = 33 K)

Yoshikazu Mizuguchi[1,2,3], Hiroyuki Takeya[1,3], Yasuna Kawasaki[1,2,3], Toshinori Ozaki[1,3], Shunsuke Tsuda[1,3], Takahide Yamaguchi[1,3] and Yoshihiko Takano[1,2,3]

1.*National Institute for Materials Science, 1-2-1 Sengen, Tsukuba 305-0047, Japan*

2.*University of Tsukuba, 1-1-1 Tennodai, Tsukuba 305-0001, Japan*

3.*JST, TRIP, 1-2-1 Sengen, Tsukuba 305-0047, Japan*

Abstract

We synthesized the new Fe-based superconductor $K_{0.8}Fe_2Se_2$ single crystals. The obtained single crystal exhibited a sharp superconducting transition, and the onset and zero-resistivity temperature were estimated to be 33 and 31.8 K, respectively. A high upper critical field of 192 T was obtained. Anisotropy of superconductivity of $K_{0.8}Fe_2Se_2$ was ~3.6. Both the high upper critical field and comparably low anisotropy are advantageous for the applications under high magnetic field.

Since the discovery of superconductivity in $LaFeAsO_{1-x}F_x$, intensive studies have been performed to find new Fe-based superconductors with higher transition temperature ($T_c$) and clarify the mechanism of high-$T_c$ superconductivity [1-7]. Now we know that the superconducting properties of the Fe-based superconductor are sensitive to the local crystal structure [8,9]. The $T_c$ of the Fe-based superconductor can be explained as a function of "anion height", which is one of the structural parameters and stands for the height of anion from the Fe-square lattice [9]. An outstanding example is pressure effect of FeSe; the $T_c$ shows a huge increase from 13 to 37 K under 4-6 GPa [10-13], and the large enhancement of $T_c$ in FeSe is strongly related to the change in the anion height. This suggests that high $T_c$ could be achieved at ambient pressure by changing the local structure through elemental substitutions and/or intercalations. Recently, K intercalations into the $Fe_2Se_2$ layer, which is associated with a structural change from 11-type (P4/nmm) to 122-type (I4/mmm), was successfully achieved, and superconducting transition around 30 K was reported [14]. Here we report transport properties of $K_{0.8}Fe_2Se_2$ single crystals under magnetic fields.

The single crystals of $K_{0.8}Fe_2Se_2$ were grown by melting an FeSe precursor with K. The FeSe precursor was prepared using Fe powder (99.9 %) and Se grains (99.999%). The starting materials with a nominal composition of Fe:Se = 1:1 were placed in an alumina crucible and sealed into an evacuated quartz tube. The tube was heated at 1100 °C for 50 h and then annealed at 380 °C for 100 h. The FeSe precursor and K grain with a nominal composition of K: FeSe = 0.8:2 were placed in an alumina crucible and sealed into an arc-welded stainless-steel tube. The sample was heated at 1030 °C for 2 h and cooled down to 750 °C with a rate of –6 °C/h. The obtained crystals were characterized by x-ray diffraction with Cu-K$\alpha$ radiation using the 2$\theta$-$\theta$ method. An actual composition of the crystal was investigated using energy dispersive x-ray spectrometry (EDX). Temperature dependence of magnetization after both zero-field cooling (ZFC) and field cooling (FC) was measured using a superconducting quantum interference device (SQUID) magnetometer with an applied field of 20 Oe. In-plane resistivity measurement was performed using the four-terminal method from 300 down to 10 K. Four terminals were attached on a fresh (as-cleaved) surface of the crystal in 10 minutes and sealed into an evacuated sample space of Physical Property Measurement System (Quantum Design). To investigate an upper critical field and anisotropy of superconductivity in $K_{0.8}Fe_2Se_2$, the temperature dependence of resistivity from 50 to 10 K was measured under magnetic fields both parallel and perpendicular to $c$-axis up to 7 T with an increment of 1 T.

We obtained large plate-like and dark-shining single crystals as displayed in a optical microscope image of the cleaved surface as shown in the inset of Fig. 1. The cleaved surface of the crystal seemed to degrade with air exposure in several minutes. Figure 1 shows the x-ray diffraction profile for the powdered crystals. The peaks were characterized using the space group of I4/mmm. The lattice constants were estimated to be $a$ = 3.9034 Å and $c$ = 14.165 Å. The obtained lattice constants are almost the same as the previous reports, but the $c$ axis seems to be comparably larger than that in Refs. 14 and 15. The actual composition of the crystal was estimated to be K:Fe:Se = 0.6:1.5:2 using an average of 4 points of the EDX measurements. The existence of deficiencies at both the K- and Fe-site is consistent with the previous result.

Figure 2 shows the temperature dependence of magnetization for the $K_{0.8}Fe_2Se_2$ crystal. A sharp drop corresponding to the superconducting transition was observed at 31 K. A large difference in magnetization curve between ZFC and FC indicates the existence of strong pinning that could be due to the K and Fe deficiency.

Figure 3 displays the temperature dependence of resistivity for the $K_{0.8}Fe_2Se_2$ crystal. At high temperatures, resistivity increases with decreasing temperature and exhibits a broad hump around 207 K. With further cooling, metallic behavior is observed and superconducting transition appears at 33 K. The onset and zero-resistivity temperature were estimated to be $T_c^{onset}$ = 33.0 and $T_c^{zero}$ = 31.8 K, respectively. The observed $T_c$ are higher than those reported in the previous report, showing $T_c^{onset}$ = 30.1 and $T_c^{zero}$ = 27.4 K [14]. Furthermore, the temperature at which the hump appeared in the previous report was around 100 K, much lower than that in our measurement. The differences in both $T_c$ and the anomaly temperature could be related to the sample quality and/or degradation of the crystals in air. In fact, the residual resistivity ratio between 207 and 33 K is a very high value of 46, which indicates that our single crystals are of high quality and/or fresh, and the observed transport properties are intrinsic. We think that the $T_c$ of $K_{0.8}Fe_2Se_2$ is correlated with the temperature of the broad hump; higher $T_c$ will appear when the broad hump appeared at a higher temperature. Furthermore, the resistivity-temperature curve exhibits a Fermi-liquid-like behavior above $T_c$, which would suggest the existence of strong electron correlation. For a greater understanding of transport properties of $K_{0.8}Fe_2Se_2$, detailed transport, structural and magnetic studies should be performed.

The temperature dependence of resistivity from 50 to 10 K under magnetic fields both parallel and perpendicular to the $c$-axis is plotted in Fig. 4 (a) and (b). A clear difference in superconductivity under fields was observed between under $B//c$ and $B//ab$. To discuss anisotropy of superconductivity precisely, $T_c^{onset}$ was defined as the temperature where the resistivity was 90 % of that just above the superconducting transition. The $T_c$ was estimated using a criterion of $\rho < 1 \times 10^{-6}$. The estimated upper critical field ($B_{c2}$) and irreversible field ($B_{irr}$) are plotted in Fig. 4 (c) and (d) as a function of temperature, respectively. The $B_{c2}^{B//c}(0)$ and $B_{c2}^{B//ab}(0)$ were estimated to be 54 and 192 T using the WHH theory [16], which gives $B_{c2}(0) = -0.69 T_c (dB_{c2}/dT)|_{T_c}$. The anisotropy, $B_{c2}^{B//c}(0K) / B_{c2}^{B//c}(0K)$ is estimated to be ~3.6. The high upper critical field and comparably low anisotropy are advantageous for applications such as a superconducting wire for the high field magnets.

In conclusion, we obtained $K_{0.8}Fe_2Se_2$ single crystal by melting K and the FeSe precursor. The obtained single crystals were dark-shining, plate-like and cleavable. We carried out the resistivity measurement using the fresh (as-cleaved) surface of the $K_{0.8}Fe_2Se_2$ crystal, and observed high $T_c^{onset}$ of 33 K. To understand the intrinsic nature of superconductivity in $K_xFe_2Se_2$, further investigation on transport, structure and magnetism is needed. The $B_{c2}^{B//ab}(0)$ is estimated to be 192 T. Due to the high upper critical field and the lower toxicity compared to As, $K_xFe_2Se_2$ could be a potential materials for the applications under high magnetic field.

Acknowledgement
This work was partly supported by Grant-in-Aid for Scientific Research (KAKENHI).

Figure captions

Fig. 1. X-ray diffraction pattern of the powdered $K_{0.8}Fe_2Se_2$. The numbers in the figure indicate the mirror index. Peaks marked by x correspond to that of PbO-type FeSe. The inset shows the optical microscope image of the cleaved surface.

Fig. 2. Temperature dependence of magnetization for the $K_{0.8}Fe_2Se_2$ crystal.

Fig. 3. Temperature dependence of resistivity for the $K_{0.8}Fe_2Se_2$ crystal.

Fig. 4. (a) and (b) show the temperature dependence of resistivity for the $K_{0.8}Fe_2Se_2$ crystal under magnetic fields of $B//c$ and $B//ab$ up to 7 T with an increment of 1 T. (c) and (d) show the temperature dependence of $B_{c2}$ and $B_{irr}$.

Fig. 1

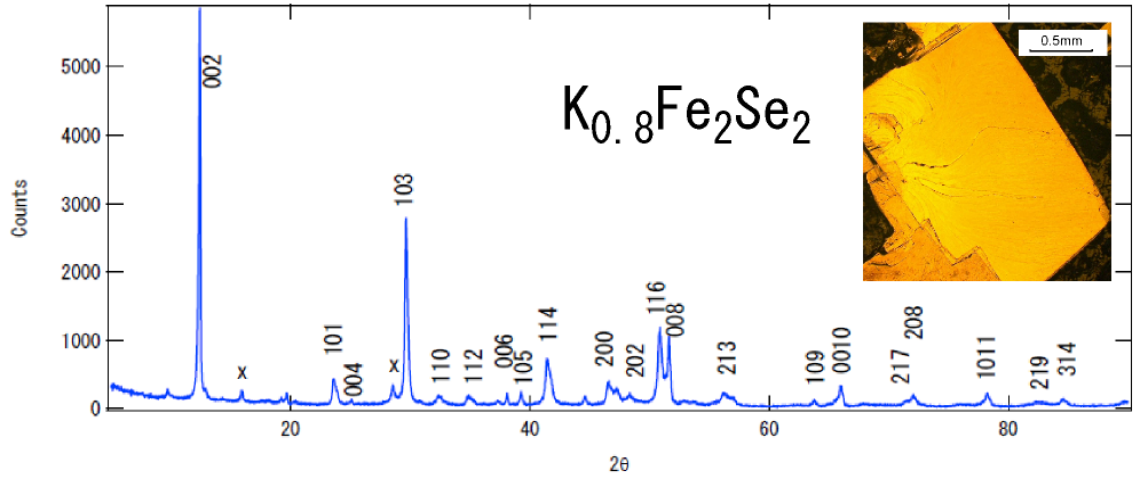

Fig. 2

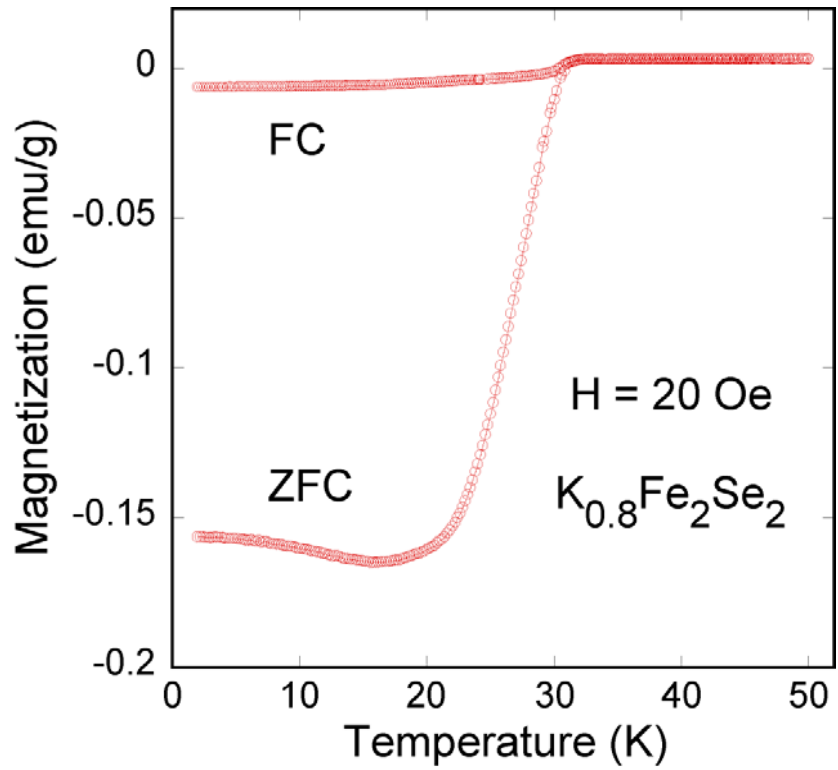

Fig. 3

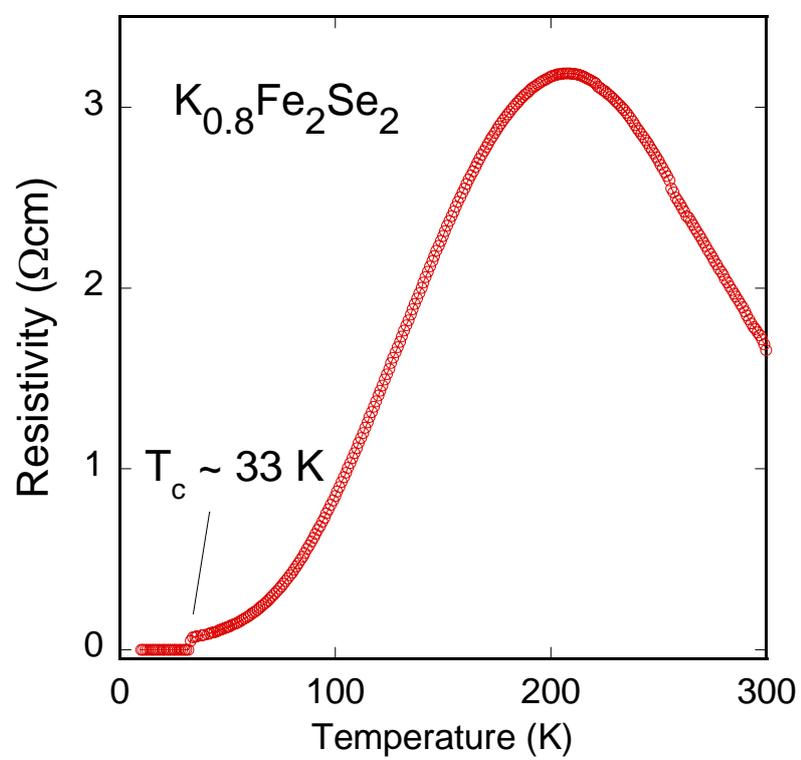

Fig. 4

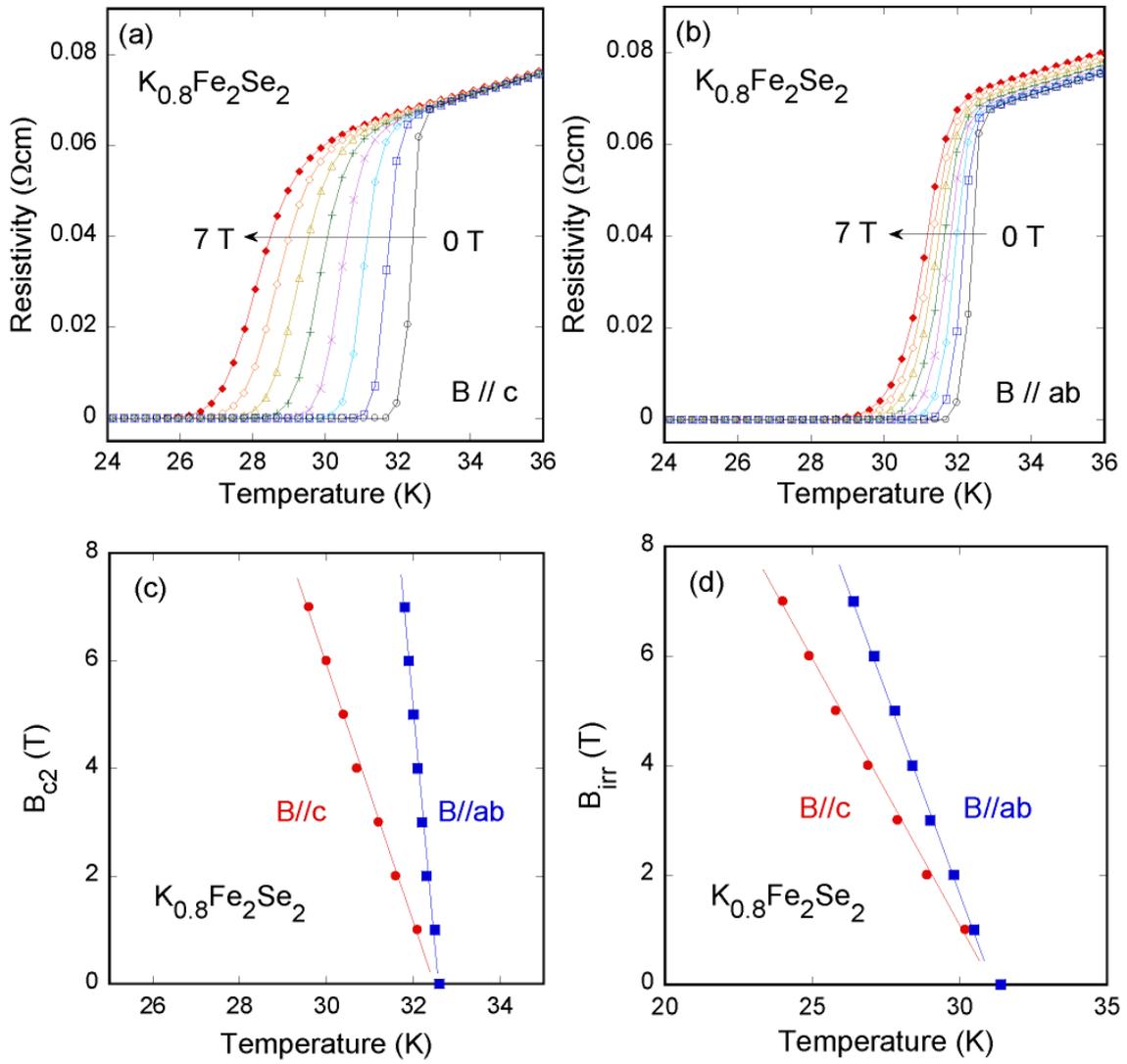

9